\documentclass{sig-alternate}

\usepackage{graphicx}
\usepackage{amsmath,amssymb} 
\usepackage{color}

\usepackage{cite}
\usepackage{multirow}
\graphicspath{{./images/}}

\newcommand{\hide}[1]{}

\newcommand{\etal}{\textit{et al.}~}

\title{Bloom Filters and Compact Hash Codes for\\ Efficient and Distributed Image Retrieval}

\author{
\alignauthor{Andrea Salvi\titlenote{andreasalvi89@gmail.com}, Simone Ercoli, Marco Bertini and Alberto Del Bimbo}\\
\affaddr{Media Integration and Communication Center, Universit\`a degli Studi di Firenze}\\
\affaddr{Viale Morgagni 65 - 50134 Firenze, Italy}\\
\email{ \{\textit{name.surname}\}@unifi.it }
}

\begin{document}






%

\date{30 July 1999}

\maketitle
\begin{abstract}
This paper presents a novel method for efficient image retrieval, based on a simple and effective hashing of CNN features and the use of an indexing structure based on Bloom filters. These filters are used as gatekeepers for the database of image features, allowing to avoid to  perform a query if the query features are not stored in the database and speeding up the query process, without affecting retrieval performance. Thanks to the limited memory requirements the system is suitable for mobile applications and distributed databases, associating each filter to a distributed portion of the database. Experimental validation has been performed on three standard image retrieval datasets, outperforming state-of-the-art hashing methods in terms of precision, while the proposed indexing method obtains a $2\times$ speedup.
\end{abstract}

%
%

%
%

%
%



\section{Introduction}\label{sec:intro}
Content based image retrieval (CBIR) has been an active research topic in computer vision and multimedia in the last decades, and it is still very relevant due to the emergence of social networks and the creation of web-scale image databases.
Most of the works have addressed the development of effective visual features, from engineered features like SIFT and GIST to, more recently, learned features such as CNNs \cite{eccv2014-babenko}. To obtain scalable CBIR systems features are typically compressed or hashed, to reduce their dimensionality and size. However, research on data structures that can efficiently index these descriptors has attracted less attention, and typically simple inverted files (e.g.~implemented as hash tables) are used.

In this paper we address the problem of approximate nearest neighbor (ANN) image retrieval proposing a simple and effective data structure that can greatly reduce the need to perform any comparison between the descriptor of the query and those of the database, when the probability of a match is very low. Considering the proverbial problem of finding a needle in a haystack, the proposed system is able to tell when the haystack probably contains no needle and thus the search can be avoided completely.

To achieve this we propose a novel variation of an effective hashing method for CNN descriptors, and use this code to perform ANN retrieval in a database. To perform an immediate rejection of a search that should not return any result we store the hash code in a Bloom filter, i.e.~ a space efficient probabilistic data structure that is used to test the presence of an element in a set. To the best of our knowledge this is the first time that this data structure has been proposed for image retrieval since, natively, it has no facility to handle approximate queries. We perform extensive experimental validation on three standard datasets, showing how the proposed hashing method improves over state-of-the-art methods, and how the data structure greatly improves computational cost and makes the system suitable for application to mobile devices and distributed image databases.

\section{Previous works}\label{sec:previous-work}

\textit{\textbf{Visual features.}}
SIFT descriptors have been successfully used for many years to perform CBIR. Features have been  aggregated using Bag-of-Visual-Words and, with improved performance, using VLAD \cite{jegouTPAMI2012} and Fisher Vectors \cite{sanchez2013image}.

The recent success of CNNs for image classification tasks has suggested their use also for image retrieval tasks. Babenko \etal\cite{eccv2014-babenko} have proposed the use of different layers of CNNs as features, compressing them with PCA to reduce their dimensionality, and obtaining results comparable with state-of-the-art approaches based on SIFT and Fisher Vectors. Aggregation of local CNN features using VLAD has been proposed in \cite{Ng-2015-CVPRW}, while Fisher Vectors computed on CNN features of objectness window proposals have been used in \cite{iccv_vsm-2015}.

\textit{\textbf{Hashing.}}
One of the most successful visual feature hashing methods presented in the literature is Product Quantization (PQ), proposed by J\'egou \etal\cite{jegou-2011}. In this method the feature space is decomposed into a Cartesian product of subspaces with lower dimensionality, that are quantized separately. The method has obtained state-of-the-art results on a large scale SIFT and GIST features dataset. 
The good performance of the Product Quantization method has led to development of several related methods that introduce variations and improvements. 
Norouzi and Fleet \cite{norouzi-2013} have built two variations of k-means (Orthogonal k-means and Cartesian k-means) upon the idea of compositionality of the PQ approach.
Ge \etal\cite{ge-2013} have improved PQ minimizing quantization distortions w.r.t.~space decomposition and quantization codebooks, in their OPQ method; He \etal\cite{he2013k} have approximated the Euclidean distance between codewords in k-means method, proposing an affinity-preserving technique.
More recently, Kalantidis and Avrithis \cite{kalantidis2014locally} have proposed to use a local optimization over a rotation and a space decomposition, applying a parametric solution that assumes a normal distribution, in their vector quantization method (LOPQ).

Most of recent approaches for CNN features hashing are based on simultaneous learning of image features and hash functions as in the method of Gao \etal\cite{Gao-2015}, that uses visual and label information to learn a relative similarity graph, to reflect more precisely the relationship among training data.

Unsupervised two steps hashing of CNN features has been proposed by Lin \etal\cite{lin2015deep}. In the first step Stacked Restricted Boltzmann Machines learn binary embedding functions, then fine tuning is performed to retain the metric properties of the original feature space.

\textit{\textbf{Indexing.}}
Typically hashed features are stored in inverted files. A few works have studied other data structures to speed up approximate nearest neighbors.
Babenko and Lempitsky \cite{babenko-2012} have proposed an efficient similarity search method that generalizes the inverted index; the method, called inverted multi-index (Multi-D-ADC), replaces vector quantization inside inverted indices with product quantization, and builds the multi-index as a multi-dimensional table. 
Ercoli \etal\cite{ercoli2015compact} have proposed an hashing method that improves over PQ by performing multiple assignments to k-means centroids, and have stored the hash codes in Marisa Tries to greatly compress their storage.

\textit{\textbf{Bloom filter.}}
Bloom filter and its many variants have received an extremely limited attention from the vision and multimedia community, so far. Inoue and Kise \cite{inoue-2009} have used Bloomier filters (i.e.~an associative array of Bloom filters) to store PCA-SIFT features of an objects dataset more efficiently than using an hash table; they perform object recognition by counting how many features stored in the filters are associated with an object. Bloom filter has been used by Danielsson \cite{Danielsson-2015} as feature descriptor for matching keypoints. Similarity of descriptors is evaluated using the ``union" operator.
Srijan and Jawahar have proposed to use Bloom filters to store compactly the descriptors of an image, and use the filter as postings of an inverted file index in \cite{Srijan-2012}.

\section{The Proposed Method}\label{sec:method}
In the proposed approach, differently from \cite{Gao-2015}, we learn a vector quantizer separately from the CNN features, so to easily replace different and pre-trained CNN networks for feature extraction, without need of retraining. Moreover, we propose to include Bloom filters into feature indexing structures to improve the speed of queries. Bloom filters act as gatekeepers that rule out immediately, with a very limited memory cost, if a query should be completely performed or if it can be avoided. The proposed data structure is very suitable for mobile and distributed applications.

\subsection{Quantization Algorithm}\label{sec:quantization-method}

The proposed approach is a variation of \cite{ercoli2015compact}, which is an efficient method for mobile visual search based on a multiple assignment k-means hashing schema (\textit{multi-k-means}) that obtained very good results, compared to PQ, on the BIGANN dataset.

\begin{figure}[htbp]
\centering%
\includegraphics[scale=0.25]{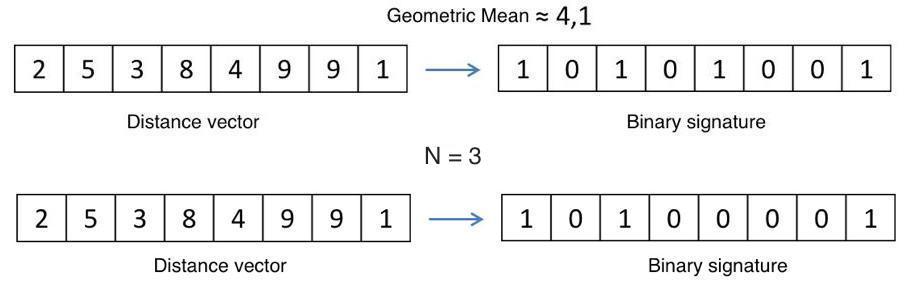}
\vspace{-2mm}
\caption{Binarization examples with a distance vector of 8 elements: (\textit{top})  geometric mean (MEAN method); (\textit{bottom}) smaller distances $N=3$ (MINx method).}\label{fig:bin-method} 
\end{figure}

The first step of the method consists in learning a standard k-means dictionary with a small number of centroids (to maintain a low computational cost). Each centroid is associated to a bit of the hash code, that has thus length equal to the number of centroids. The bit is set to 1 if the feature is assigned to the centroid, 0 otherwise. A feature can be assigned to more than one centroid, and it is assigned to it if the distance from the centroid is less than the mean distance from all the centroids (Figure \ref{fig:bin-method}, top). Instead, in this work we select a fixed number $N$ of distances and we set to 1 all the bits associated to the smaller $N$ distances (Figure \ref{fig:bin-method}, bottom). In the following we refer to this method as MINx.
This change has proven to be more efficient when coding CNN feature descriptors, that were used in the experiments.

Approximate nearest neighbor retrieval of image descriptors is performed in two steps: in the first step is performed an exhaustive search over the binary codes using Hamming distances, to reduce negative effects of quantization errors. All the binary codes with Hamming distance below a threshold are selected. In the second step the candidate neighbors are ranked according to the distance computed using the full feature vector using \textit{cosine distance}, that proved to be more effective than $L_2$ during the experiments.

\hide{
\begin{equation}\label{eq:dist}
d_{cos}(p,q):= \cos(p,q) = \dfrac{p \cdot q}{||p||\ ||q||}=\dfrac{\sum_{i}{p_iq_i}}{\sqrt{\sum_{i}{p_i}}\sqrt{\sum_{i}{q_i}}}
\end{equation}
}

\subsection{Bloom Filter Algorithm}\label{sec:bloom-method}
\begin{figure}[htbp]
\centering%
\includegraphics[scale=0.24]{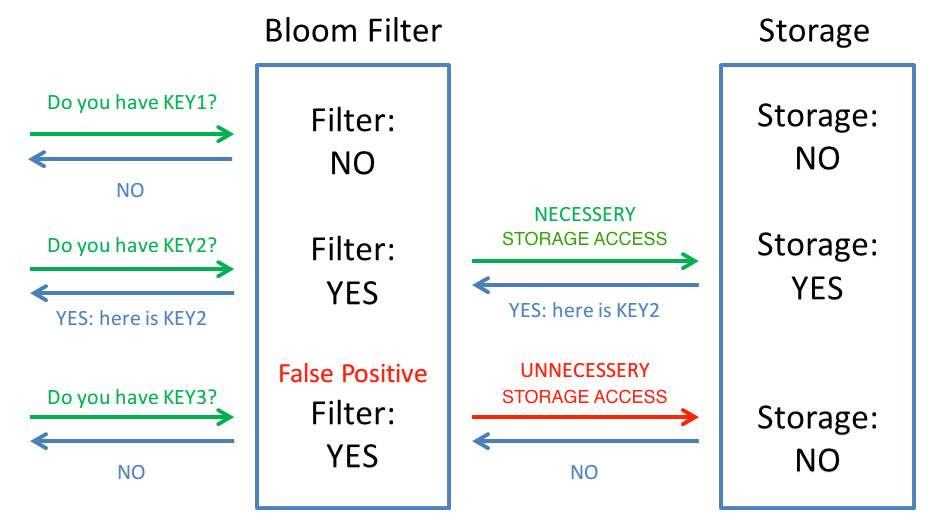}
\vspace{-2mm}
\caption{Memory accesses with Bloom filters}\label{fig:bloom_filters}\end{figure}

To improve search of feature vectors we also introduce the use of \textit{Bloom filters} \cite{bloom1970space}. Typically this type of structures are used to speed up the answers in a key-based storage system (Figure \ref{fig:bloom_filters}).

A Bloom filter is an efficient probabilistic data structure used to test if an element belongs to a set or not. This structure works with binary signatures, and can provide false positive response but not false negative and more elements are inserted into the structure and more high is the probability to obtain a false positive. To insert an element inside a Bloom filter we need to define $k$ hash functions which locate $k$ positions inside the array, setting them to 1.

To check the presence of an element inside a Bloom filter we to compute the $k$ hash functions over the element and check the related positions inside the array. If just one bit of these positions is equal to 0 it means that the element is not present inside the array; if all the checked bits are equal to 1 it means that either the element is inside the array or we have a false positive. We used the method of \cite{Kirsch-2008} to create the $k$ functions from just two hash functions.

\begin{figure}[htbp]
\centering%
\includegraphics[scale=0.18]{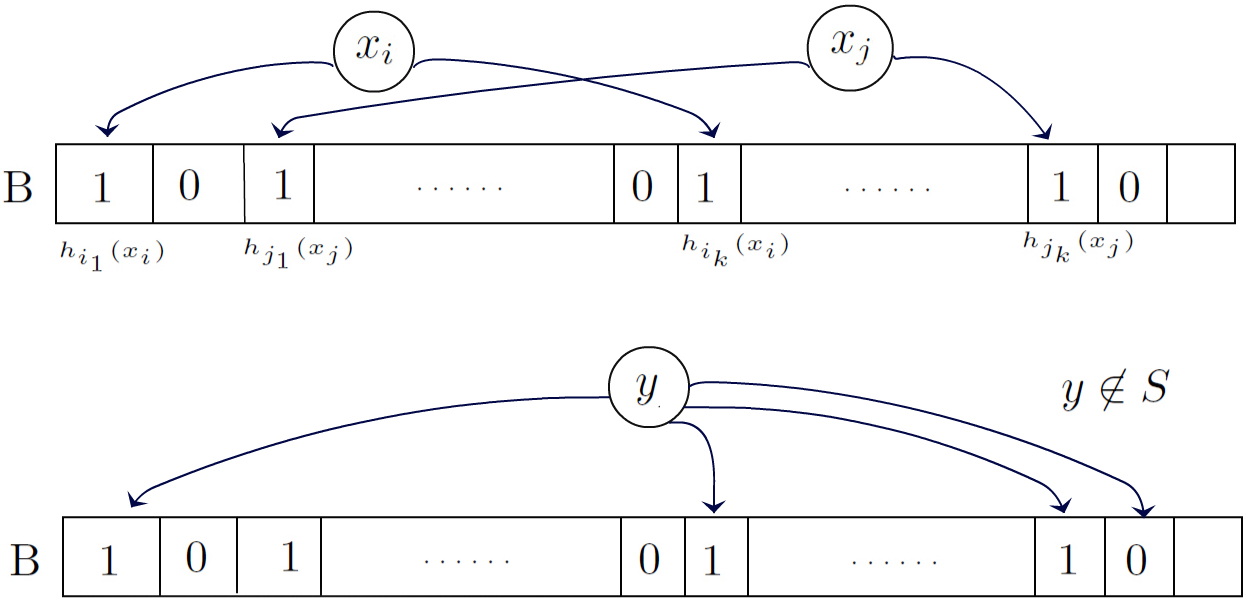}
\vspace{-3mm}
\caption{Example Bloom filter: (\textit{top}) Insertion, (\textit{bottom}) Search}\label{fig:bloom_filter}
\end{figure}

A useful property of Bloom filter is that we can measure the presence of a false positive with probability:
\vspace{-4pt}
\begin{equation}
(1-e^{-kn/m})^{k}=(1-p)^{k}=\epsilon
\label{probabilitafp}
\end{equation}

where $m$ is the bit number of the array, $n$ is the number of inserted items, $p$ is the probability that one position of the array is equal to 0, and $k$ is the number of hash functions. We can obtain the optimal value $k$ which minimizes false positive probability:
\vspace{-4pt}\vspace{-4pt}
\begin{equation}
\tilde{k}=\ln 2(m/n).
\label{kottimo}\end{equation}

Supposed that $p=0.5$ we can write out $\epsilon$ like
\vspace{-4pt}
\begin{equation}
\epsilon=0.5^{\tilde{k}}=(0.6185)^{m/n}
\label{eps}\end{equation}

So $n$ is strictly related to $m$, and in general $m=O(n)$ it is a good compromise.

Storing in the Bloom filter hash codes that are designed for ANN, as those of Sect.~\ref{sec:quantization-method}, results in a data structure that is similar, from a practical point of view, to distance-sensitive Bloom filters proposed in \cite{kirsch-2006}, where LSH functions are used as $k$ hash functions.

\hide{
\begin{table}[htbp]
\centering
\caption{
        Bloom Filter false positive probability related to $m$ 
	}
\resizebox{0.3\columnwidth}{!}{
\begin{tabular}{ | r | r | r |}
\hline
$m$ & $\epsilon$ & \% \\
\hline
n & 0.61 & 61\% \\
2n & 0.38 & 38\% \\
5n & 0.09 & 9\% \\
10n & 0.008 & 0.8\% \\
\hline
\end{tabular}
}

\label{tab:probFP}
\renewcommand\arraystretch{1}

\end{table}
}

\subsection{Retrieval System}\label{sec:retrieval-system}

\begin{figure}[htbp]
\centering%
\includegraphics[scale=0.18]{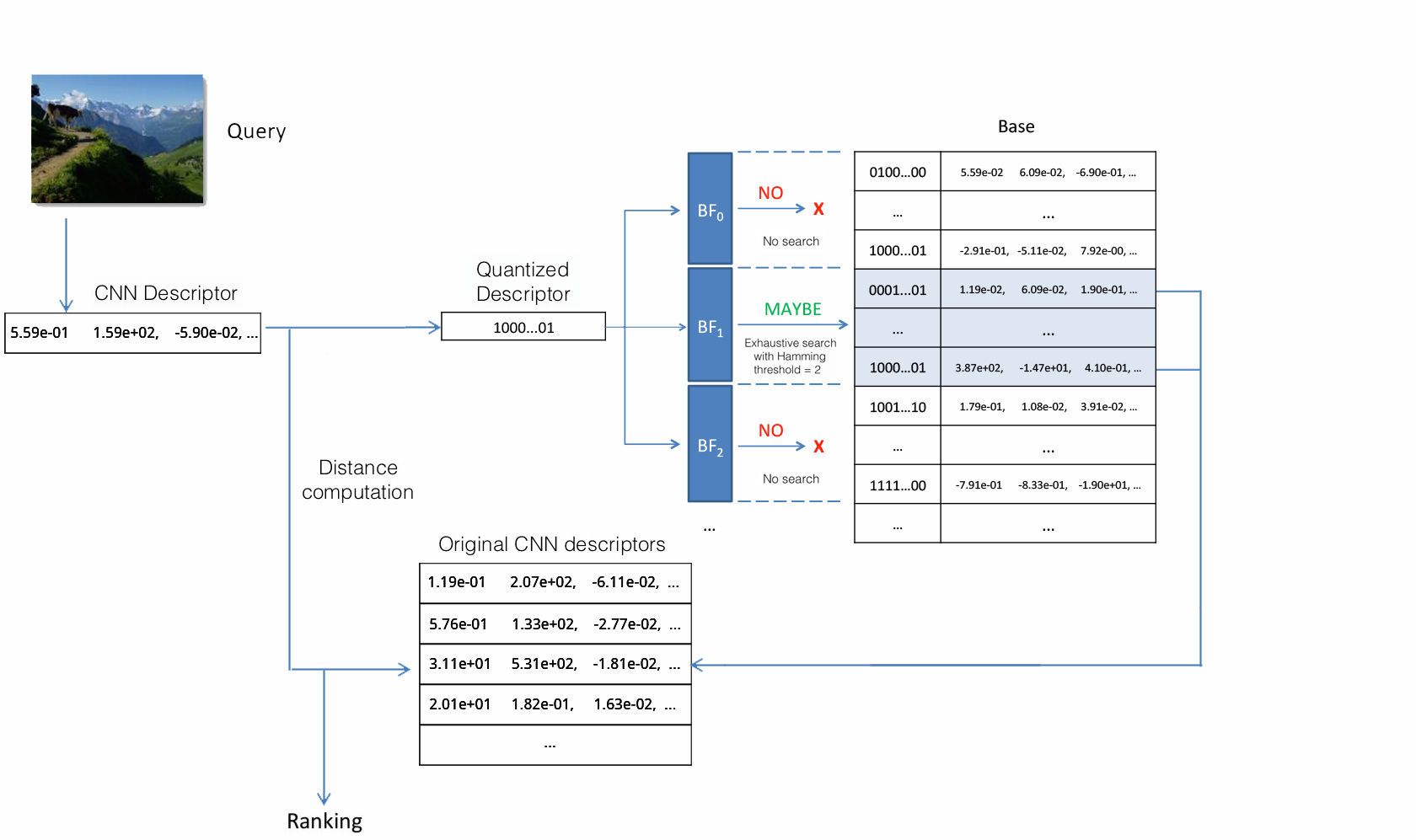}
\vspace{-4mm}
\caption{System overview.}\label{fig:system}
\end{figure}

Our proposed retrieval system merges the methods introduced in \ref{sec:quantization-method} and \ref{sec:bloom-method}. Regarding visual feature hashing we have applied the proposed method to CNNs features.
Our system (Figure \ref{fig:system}) provides a initial phase were descriptors are extracted from base images, binarized following one of the methods introduced in \ref{sec:quantization-method} and saved inside a data structure composed by a set of inverted files of hashes implementing an horizontal partition of data (allowing to distribute the database as ``shard"), each one guarded by a Bloom filter. The hash code is also added to the Bloom filter of the corresponding inverted file.

During the search phase we extract the CNN descriptor from query images, compute the hash code,  and check the presence of the hash in the Bloom filters, each of which guard a subset of the base. If one of this Bloom filters gives a positive response (this means that we have a positive or a false positive match), all the hash codes within an Hamming distance threshold are used to select the full feature vector. This provides a great speedup in the approximate nearest neighbor retrieval since we consider only descriptors from base coded by a Bloom filter, and below the Hamming threshold value. For each resulting original CNN descriptor we compute the distance and we rearrange results to obtain a ranked list of vectors.

\section{Experimental results}\label{sec:experiments}

\subsection{Datasets and Configurations}\label{sec:dataset}

We tested our system using three standard dataset: INRIA Holidays \cite{Inria}, Oxford 5K \cite{Oxford} and Paris 6K \cite{Paris}. We used the query images and ground truth provided for each dataset, adding 100,000 distractor images from \textit{Flickr 100} \cite{Philbin07}. When testing on a dataset training is performed using the other two datasets. Features have been hashed to 64 bits binary codes, a length that has proved to be the best compromise between compactness and representativeness. 
Other parameters used for hashing were:

\noindent
-- number of $N$ nearest distances used in the hash code computation ($N \in \{6,10,16,32,40\}$);

\noindent
-- Hamming distance threshold $\in \{2,4,6,10,16,30\}$.

For the sake of brevity, in the following we report only the best combinations.
For the evaluation we used the Mean Average Precision (MAP) metric.
\hide{
Given the \emph{Precision at k} value

\begin{equation}
P@k:=\dfrac{\#imgRelevant\ with\ rank\ \leq k}{k}
\end{equation}.

which is the ratio between the relevant images in the first k position and k the number of retrieved images, we can define the \emph{Average Precision} value for a single query $q$

\begin{equation}
AP(q):=\dfrac{1}{R}\sum_{i=1}^R P@k_i
\end{equation}

and compute the mean over $Q$ queries with:

\begin{equation}
MAP(Q):=\dfrac{1}{|Q|}\sum_{j=1}^{|Q|} AP(q_j)
\end{equation}
} 
The CNN features used in the following experiments have been extracted using the 1024d average pooling layer of GoogLeNet \cite{Szegedy-2015}, that in initial experiments has proven to be more effective than the FC7 layer of VGG \cite{Simonyan-2014} used in \cite{iccv_vsm-2015}.

\subsection{Results}\label{sec:results}
\begin{figure}[htbp]
\centering
\includegraphics[width=1\columnwidth]{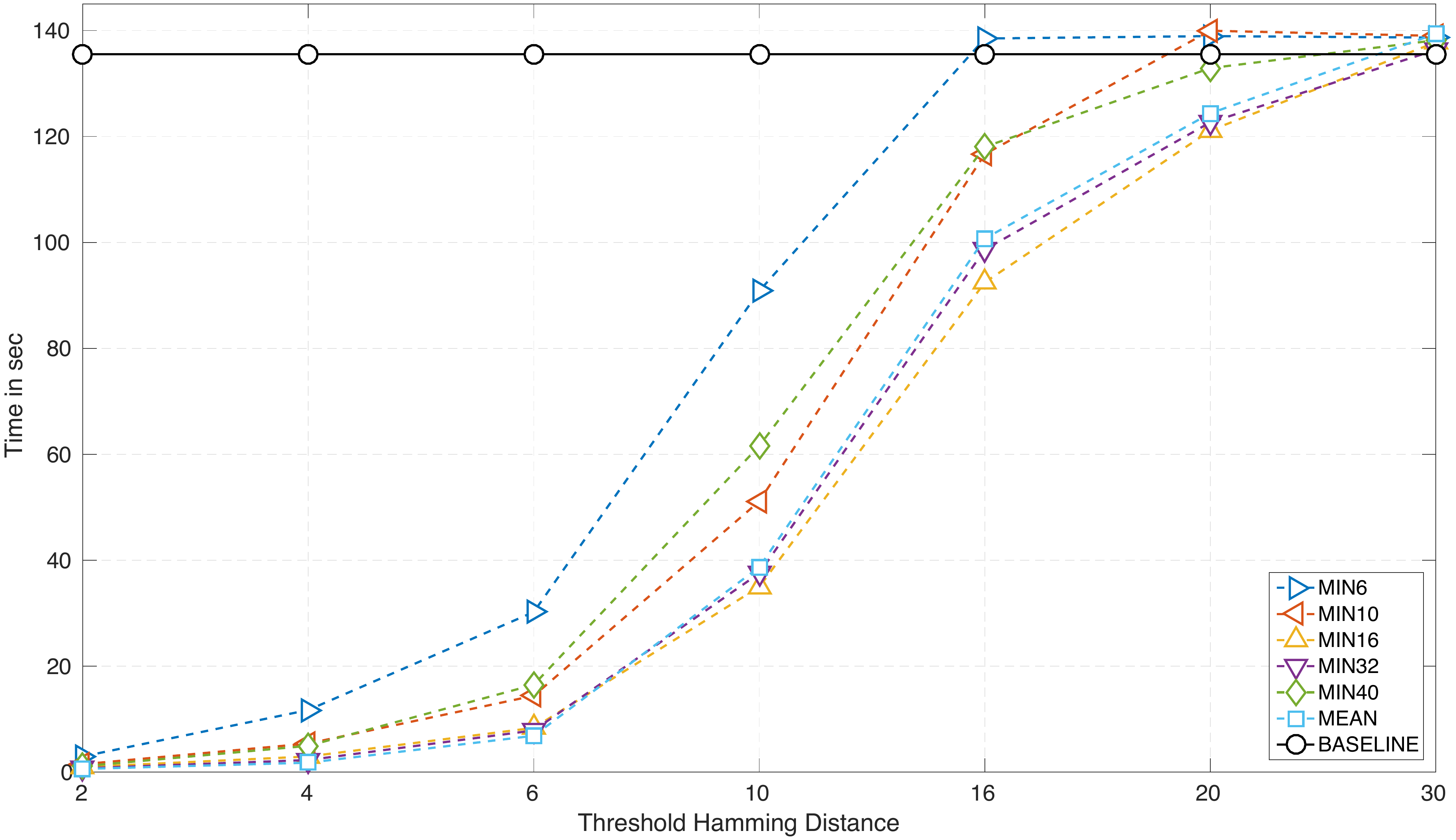}
\vspace{-4mm}
\caption{INRIA Holidays: time comparison for the MINx method, the MEAN \cite{ercoli2015compact} method and baseline without hashing.}\label{fig:inria-time}
\end{figure}

In the first experiment we evaluate the effects of the method parameters, comparing the proposed hashing approach (MINx) with the original method of \cite{ercoli2015compact} (MEAN), a baseline that uses no hashing, and several state-of-the-art methods, among which the recent UTH method \cite{lin-2015}. The best combinations of MINx are reported, compared on the three  datasets in terms of MAP. As expected the uncompressed features perform better, but the MIN6 setup, with an Hamming distance $\ge 6$ has comparable results, and greatly outperforms any state-of-the-art hashing method.
Time results in seconds, for INRIA Holidays dataset, are reported in Fig.~\ref{fig:inria-time}. A $2\times - 10\times$ speedup can be obtained with Hamming distances between 6 and 10. Similar results, not reported here for the sake of brevity, have been obtained on Oxford 5K and Paris 6K datasets. 
\vspace{-4pt}

\begin{table}[!ht]  
  \centering
  \caption{MAP results on Holidays, Oxford 5K and Paris 6K datasets. The proposed MINx method outperforms all the current state-of-the-art methods. All hashes are 64 bit long.}\label{tab:comparisons}
\resizebox{0.8 \columnwidth}{!}{
  \begin{tabular}{|c|c|c|c|}
    \hline
    \textbf{Method} & \textbf{Holidays}  & \textbf{Oxford 5K} & \textbf{Paris 6K} \\ \hline \hline
    ITQ \cite{gong-2011}	&	53.68	&	23.00	&	 - \\ \hline
    BPBC \cite{gong2013learning} 	&	38.10	&	22.51 	& - \\ \hline
    PCAHash \cite{gong-2011} 	&	52.80	&	23.90	& - \\ \hline
    LSH \cite{datar2004locality} 	&	43.08	&	23.91	&	 - \\ \hline
    SKLSH \cite{raginsky2009locality} 	&	24.09	&	13.39	&	 - \\ \hline
    SH \cite{weiss-2009} 	&	52.22	&	23.24	&	 -\\ \hline
    SRBM \cite{chandrasekhar2015compact}	&	51.58	&	21.23	&	 -\\ \hline
    UTH \cite{lin-2015} 	&	57.10 	&	24.00	&	 -\\ \hline \hline
    \textbf{\begin{tabular}[c]{@{}c@{}}MIN6\\ Thr. 10\end{tabular}} 	&	\textbf{75.62}		&	\textbf{46.03}	&	 67.57\\ \hline
    \textbf{\begin{tabular}[c]{@{}c@{}}MIN6\\ Thr. 16\end{tabular}} 	&	\textbf{75.63}	&	\textbf{46.06}	&	 \textbf{67.84}\\ \hline
    \textbf{\begin{tabular}[c]{@{}c@{}}MIN10\\ Thr. 10\end{tabular}} 	&	74.33		&	45.66	&	 60.84 \\ \hline
    \textbf{\begin{tabular}[c]{@{}c@{}}MIN10\\ Thr. 16\end{tabular}} 	&	\textbf{75.63}		&	46.04	&	 67.70 \\ \hline
    \textbf{\begin{tabular}[c]{@{}c@{}}MEAN\cite{ercoli2015compact}\\ Thr. 10 \end{tabular}} & 	68.26	&		36.22	&	48.02  \\ \hline
    \textbf{\begin{tabular}[c]{@{}c@{}}Baseline\end{tabular}} &	75.63	&	46.06	&	 67.84 \\ \hline
  \end{tabular}
}
\end{table}

In the second experiment we evaluate a use case in which a database of images is queried with a large number of images that do not belong to it. Hash codes have been computed with different variants of the proposed hashing method. The database contains the Paris 6K images, and it is queried with all the query images of Paris 6K and all the 100,000 distractor images. A different number of Bloom filters, with different sizes is tested and compared against a baseline that does not use any Bloom filter. MAP values and query time in seconds are reported in Tab.~\ref{tab:map-bf-paris6k} (MAP in the first row and time in the second). The speedup obtained is about $2 \times$ since a large number of distractor queries are immediately stopped by the system; the slight increase in MAP is due to the beneficial effect of elimination of some false positives of the Paris 6K images, that do not result in retrieving wrong dataset images.

\begin{table}[!htb]
\centering
\caption{MAP+time (secs.) obtained on Paris 6K with the proposed system with different numbers of Bloom filters (1, 2 and 5) and with a baseline without filters. $2n$ and $5n$ are the size of the filters, where $n$ is the number of stored elements. \textit{Thr.} is the maximum Hamming distance used for hash code retrieval.}\label{tab:map-bf-paris6k}
\resizebox{1 \columnwidth}{!}{
\begin{tabular}{|c|c|c|c|c|c|c|c|}
\hline
	& \multirow{2}{*}{\textbf{\begin{tabular}[c]{@{}c@{}}No\\ BF\end{tabular}}} & \multicolumn{2}{c|}{\textbf{1 BF}} & \multicolumn{2}{c|}{\textbf{2 BF}} & \multicolumn{2}{c|}{\textbf{5 BF}} \\ \cline{3-8} 
	&	& \textbf{2n}	&	\textbf{5n}	&	\textbf{2n}	&	\textbf{5n}	&	\textbf{2n}	&	\textbf{5n}	\\ \hline
\textbf{MIN6}	& 67.57		& 67.57     &	67.57	& 67.94		& \textbf{70.96}	& 67.53	& 68.21          \\ 
\textbf{Thr. 10}& 460.31		& 242.72		& 173.46		& 311.26		& \textbf{205.08}	& 382.14	& 366.61	\\ \hline

\textbf{MIN10}	&	67.69	& 67.69	&	67.69	&	68.37	&	68.46	&	67.66	&	66.70	\\
\textbf{Thr. 16}&	553.79	& 307.46	&	174.53	&	459.74	&	272.31	&	543.65	&	430.06	\\ \hline
\end{tabular}
}
\end{table}

In the third experiment we evaluate a more challenging and large scale experiment: three datasets composed by distractor images and Holidays, Paris 6K and Oxford 5K images are built and stored in the proposed data structure. The standard dataset query images are then used to query the system. In this case we have used 10 filters to ``shard" the database that, thus,  can be distributed. Tab.~\ref{tab:map-bf-needle} reports the results in terms of MAP and time (secs.). For the sake of space we report only results for MIN6 and Hamming threshold 10. Using the proposed method results in speed improvement of $2 \times$ while improving MAP, except the Holidays dataset that only improves speed. The size of each Bloom filter is $\sim 6-62$ KB, allowing the use of the method in a mobile environment, by distributing the Bloom filters to the mobile devices and maintaining the shards of the database on the backend.
\vspace{-4pt}

\begin{table}[!htb]
\centering
\caption{MAP+time (secs.) obtained on Paris 6K, Oxford 5K and Holidays with the proposed system with 10 Bloom filters of varying size and with a baseline without filters. The database contains 100,000 distractor + the database images of each dataset.}\label{tab:map-bf-needle}
\resizebox{0.8 \columnwidth}{!}{
\begin{tabular}{|c|r|r|r|}
\hline
\textbf{\# BF}	&	\textbf{Paris 6K	}	&	\textbf{Oxford 5K}	&	\textbf{Holidays}		\\ \hline \hline
No BF	&	58.52		&	42.05		&	\textbf{59.56}		\\
		&	3.35		&	2.35			&	56.71		\\ \hline
10 BF	&	59.44		&	41.45		&	52.15		\\
10n		&	2.66		&	1.95			&	47.21		\\	\hline
10 BF	&	61.21		&	42.01		&	42.36		\\
20n		&	1.93		&	1.53			&	36.36		\\	\hline
10 BF	&	62.82		&	\textbf{42.29}		&	39.26		\\
30n		&	1.43		&	1.25			& 	31.37		\\	\hline
10 BF	&	\textbf{63.52}		&	\textbf{42.24}	 &	34.29		\\
50n		&	\textbf{1.21}		&	\textbf{1.11}			&	\textbf{26.59}		\\	\hline
\end{tabular}
}
\end{table}

\section{Conclusions}\label{sec:conclusions}
In this paper we have presented a simple and effective method for CNN feature hashing that outperforms current state-of-the-art methods on standard datasets. A novel indexing structure, where Bloom filters are used as gatekeepers to inverted files storing the hash codes, results in a $2 \times$ speedup for ANN, without loss in MAP.

\hide{
\section{Acknowledgments} 
This work is partially supported by the ``Social Museum and Smart Tourism'' project (CTN01\_00034\_231545).

This research is based upon work supported [in part] by the Office of the Director of National Intelligence (ODNI), Intelligence Advanced Research Projects Activity (IARPA), via IARPA contract number 2014-14071600011. The views and conclusions contained herein are those of the authors and should not be interpreted as necessarily representing the official policies or endorsements, either expressed or implied, of ODNI, IARPA, or the U.S. Government.  The U.S. Government is authorized to reproduce and distribute reprints for Governmental purpose notwithstanding any copyright annotation thereon.
}

\newpage
%
\bibliographystyle{abbrv}

%
%

\end{document}